\documentclass[amssymb,prb,preprint,aps]{revtex4}
\usepackage{epsfig}

\begin{document}
\title{Nonlinear effects in tunnelling escape in $N$-body quantum systems}
\author{V. Fleurov\footnote{Email:fleurov@post.tau.ac.il.}}
\affiliation{Raymond and Beverly Sackler Faculty of Exact Sciences,
School of Physics and Astronomy, \\ Tel-Aviv University, Tel-Aviv
69978 Israel. }
\author{A. Soffer}
\affiliation{Department of Mathematics, Rutgers University, New
Brunswick, NJ 08903,USA}
\begin{abstract}
We consider the problem of tunneling escape of particles from a
multiparticle system confined within a potential trap. The process
is nonlinear due to the interparticle interaction. Using the
hydrodynamic representation for the quantum equations of the
multiparticle system we find the tunneling rate and time
evolutions of the number of trapped particles for different
nonlinearity values.
\end{abstract}

\maketitle

The recent advances in the experiments on real Bose --
Einstein condensates (BEC) \cite{aemw95,dmaddkk95,bsth95} and
nonlinear optical waves\cite{fsec03} have generated a huge body of
works on the theoretical side. The standard approach on the theory
side was the use of the nonlinear Schr\"odinger equation, with a
potential term depending on the particle density (also known as
the time dependent Gross-Pitaevskii 
equation\cite{p61,g63,iasmsk98,ss99,l9800}). However, as it became
clear early in the investigation, the time behavior of such
equations was very complex and rich. The phenomena of
coherence\cite{rp02}, macroscopic tunneling\cite{sspkpl04}, vortex
formation,\cite{wh99,mahhwc99,mcwd99,igrcgglpk01} instabilities,
focusing and blowup are all new concepts which are related to the
{\em nonlinear} nature of the systems. Most of the analysis of these
hard and fundamental analytic problems are so far being dealt with
by a combination of numerical schemes (e.g. Ref.
\onlinecite{sspkpl04}) and finite dimensional phenomenological
models. The quest for a theory that can adequately give the
relevant nonlinear effects in the time dependent regime is therefore
of major current interest.

In this letter we analyze the problem of resonance and tunneling
of N-body Quantum systems by solving the corresponding nonlinear
problem in the leading relevant approximation. Our approach
combines ideas from Many body theory, nonlinear partial
differential equations and resonance theory in Quantum Mechanics
to offer a new unified approach to finding tunneling times for
both linear and nonlinear systems. We focus on the problem of a
BEC droplet in a potential well, that can tunnel through a finite
barrier from the trap. We find the leading nonlinear corrections
to the tunneling rate.

A droplet of a large number of atoms, $N$, with boson statistics
is confined by an external potential $V_{ext}({\bf R})$. The
multiparticle wave function describing the quantum state of such a
system satisfies the Schr\"odinger equation
\begin{equation}\label{1}
i\hbar\frac{\partial \Psi(\{{\bf R}_i\})}{\partial t} =
 \left(-\frac{\hbar^2}{2m}
\sum_i \frac{\partial^2}{\partial {\bf R}_i^2} + \sum_i V_{ext}({\bf R}_i)
+ \frac{1}{2} \sum_{i \neq j} V_{int}({\bf R}_i - {\bf R}_j)  \right)
\Psi(\{{\bf R}_i\})
\end{equation}
Starting from this equation we may derive equations of motion for
moments of the single particle Wigner function $\rho_W({\bf
p},\varepsilon, {\bf R},t)$.\cite{foot} We concentrate here on the condensate
behavior, which is supposed to be separated from the dissipative
excitations (a detailed discussion of this separation can be found
in Ref. \onlinecite{itg02}). The derivation can be made gauge
invariant and its details are presented in the papers \cite{lf}
where many relevant references can be also found. The most
important of the moments are the density distribution of the
particles
$$
\rho({\bf R}, t) = \frac{1}{(2\pi\hbar)^4}\int d^3pd\varepsilon
\rho_W({\bf p},\varepsilon, {\bf R},t)
$$
and the velocity field
$$
{\bf v}({\bf R}, t) = \frac{1}{(2\pi\hbar)^4} \frac{1}{\rho({\bf R}, t)}
\int d^3pd\varepsilon
{\bf p}\rho_W({\bf p},\varepsilon, {\bf R},t)
$$
for which the continuity equation
\begin{equation}\label{2}
\frac{\partial \rho({\bf R}, t)}{\partial t} + \nabla({\bf v({\bf R}, t)}
\rho({\bf R}, t)) = 0
\end{equation}
and the Euler-type equation
\begin{equation}\label{3}
\frac{\partial  {\bf v({\bf R}, t)}}{\partial t} + ({\bf v({\bf
R}, t)} \cdot \nabla) {\bf v({\bf R}, t)} = - \frac{1}{m} \nabla
(V_{ext}({\bf R}) + V_{qu}({\bf R}) + V_{eff}({\bf R}) ).
\end{equation}
are obtained. The velocity field is not necessarily potential and
three dimensional cases allows for vortices. Here the "quantum
potential" $ V_{qu}({\bf R}) = - \displaystyle \frac{\hbar^2}{2m}
\frac{\nabla^2 \sqrt{\rho({\bf R}, t)} }{\sqrt{ \rho({\bf R},
t)}}$ accounts for the quantum character of the liquid in the
droplet, whereas $ V_{eff}({\bf R})$ is the contribution due to
the inter-particle interaction. In many cases, the interaction
potential $ V_{eff}({\bf R})$ can be represented as a functional
of the particle density $\rho({\bf R}, t)$. For example, if we
assume that the inter-particle interaction is point-like, $
V_{int}({\bf R}_i - {\bf R}_j) = \frac{1}{2} \lambda \delta({\bf R}_i - 
{\bf R}_j) $ then the direct calculations, whose many aspects are
parallel to those presented in Ref. \cite{itg02}, result in the
lowest order in $\lambda$ in $V_{eff}({\bf R}) = \lambda \rho({\bf
R}, t)$. Then equations (\ref{2}) and (\ref{3}) become equivalent
to the time dependent Gross-Pitaevskii equation. Higher order term
in $\lambda$ (to be discussed separately) will produce a
correction to the above effective potential $V_{eff}({\bf R})$ as
well as a dissipative contribution due to the interaction with the
excitations above the condensate.

\begin{figure}
\includegraphics[width=9cm,angle=-90]{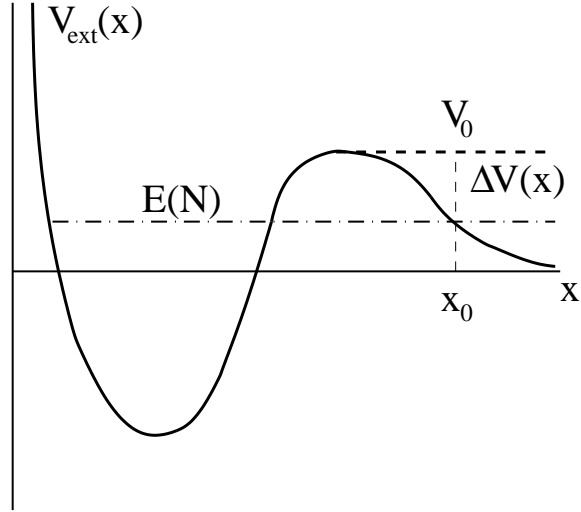}
\caption{Potential well keeping the droplet. The dashed line shows
the auxiliary potential $V_0(x)$ at larger $x$. The dash dotted line 
corresponds to the energy $E(N)$ for a given number of particles $N$.}
\label{f.1}
\end{figure}

In this presentation we will limit ourselves to discussion of the
problem of tunneling evaporation of the droplet kept within a
one-dimensional potential depicted in figure \ref{f.1}.
It means that we will have to consider the one dimensional versions of
equations (\ref{2}) and (\ref{3}), which read now
\begin{equation}\label{4}
\rho_t(x,t) + \frac{\partial}{\partial x} [\rho(x,t)v(x,t)] = 0.
\end{equation}
and
\begin{equation}\label{5}
v_t(x,t) + v(x,t) v_x(x,t) = \frac{1}{m} \frac{\partial}{\partial
x}\left [V_{ext}(x) - \frac{1}{\sqrt{\rho(x,t)}} \frac{\hbar^2}{2
m}\sqrt{\rho(x,t)} + U_{eff}(\rho(x,t))\right].
\end{equation}

In order to solve the problem we first introduce an auxiliary
confining potential $V_{0}(x)$, which coincides with the real
potential $V_{ext}(x)$ for small $x$ but differs from it (dashed
line in fig. \ref{f.1}) at large $x$. Stationary states are
possible in such potential with the density distribution following
from equation (\ref{5}) at $v(x,t)=0$ and with $V_{0}(x)$
substituted for $V_{ext}(x) $.
\begin{equation}\label{6}
V_0(x)\sqrt{\rho(x,t)} -  \frac{\hbar^2}{2
m}\frac{\partial^2}{\partial x^2}\sqrt{\rho(x,t)} +
U_{eff}(\rho(x,t))\sqrt{\rho(x,t)}
 = E(N)\sqrt{\rho(x,t)}
\end{equation}
$E(N)$ is the single particle eigen-energy in a droplet with $N$
particles. One readily recognizes a stationary Schr\"odinger-type
equation for the single particle wave function $\psi(x,t) =
\sqrt{\rho(x,t)}$ and with the effective potential
$U_{eff}(\rho(x,t))$ accounting for the many body effects. In case
of a contact interaction between the particles equation (\ref{6})
becomes celebrated stationary Gross-Pitaevskii equation.

The solution of equation (\ref{6}) at a given initial number of
particles $N_0$ is considered as an initial state of a droplet in
the real external potential $V_{ext}(x)$. In this case the density
distribution $\rho(x,t)$ becomes time dependent and its dynamics
is governed by equation (\ref{5}). Here we will follow the time
evolution of the number of particles $N(t)$ within the droplet,
defined as $N(t) = \int_0^{x_0}dx\rho(x,t) $ and with $x_0$ chosen
as the exit point of the potential $V_{ext}(x)$ at a given energy
$E(N)$ (see fig. \ref{f.1}). $N(t)$ at $t=0$ nearly coincides with
the initial number of particle $N_0$ since only exponentially
small part of the particle density remains outside the potential
well at $x > x_0$.

Equation (\ref{5}) can be solved if we assume that the velocity
field and the density distribution change adiabatically slow with
time. The applicability of this approximation will be discussed
below. Then we may neglect the time derivative $v_t(x,t) = 0$ in the
rhs of (\ref{5}). Using the fact that initially the density
distribution $\rho(x,t)$ satisfies equation (\ref{6}) we may find
the velocity field $v(x,t) = \sqrt{\frac{2}{m}\Delta V(x)}$ where
$\Delta V(x) = V_0(x) - V_{ext}(x)$

Now we find the number of particles within the well $V_{ext}(x)$
by integrating the continuity equation (\ref{4}) from 0 to the
exit point $x_0$. According to our approximation both the density
and the velocity fields vary slowly with time only due to the time
variation of the number of particles $N(t)$. The exit point $x_0$
is connected with eigen-energy $E(N)$ of the tunneling particles,
so that the velocity becomes $v(x_0,t) = \sqrt{\frac{2}{m} [V_0 -
E(N(t))]}$ and we get equation
\begin{equation}\label{9}
t= - \int_{N_0}^{N}
d\tilde N\frac{1}{\rho(x_0)\sqrt{ \displaystyle
\frac{2}{m} [V_0 - E(\tilde N)]}}.
\end{equation}
which determines implicitly the time dependence of the number of
particle $N(t)$.

In order to estimate the particle density $\rho(x_0)$ at the exit
point the particle density at the eigen energy $E(N)$ is written
as $\rho(x;N) = N \varrho(x;N)$, with $\varrho(x;N)$ normalized to
one. $\rho(x,N)$ satisfies Schr\"odinger equation (\ref{6}) but we
need to know its value only at the exit point where it is
exponentially small and we may neglect the inter-particle
interaction. As a result we get $\sqrt{\varrho(x_0;N(t))}\propto
e^{- \alpha x_0}$ with $\alpha = \frac{1}{\hbar}\sqrt{2m(V_0 -
E(N))}$.

If the dependence of the energy $E(N)$ on the number of particles
in the droplet is known, equation (\ref{9}) can be solved and the
time dependence $N(t)$ of the number of particles can be found. As
an example we carry out the calculation for the contact
inter-particle interaction when equations (\ref{4}) and (\ref{5})
correspond to the Gross-Pitaevskii equation. At not too large $N$
we may approximately assume a linear dependence of the energy on
the number of particles $E(N) = E(0) - \widetilde \lambda N$
\cite{sw} and represent the integral (\ref{9}) in the form
\begin{equation}\label{10}
\frac{1}{2} \Gamma_0 t= - \int_{y_0}^y dy \frac{e^{
\frac{1}{2} \alpha_0 x_0 y^2}}{y
\sqrt{1 + y^2 }}
\end{equation}
where $ y^2 = 2 \frac{N}{N_{GP}}, \ \ \ N_{GP} = \frac{\hbar^2
\alpha_0^2}{\widetilde \lambda m},\ \ \ \alpha_0 = \frac{1}{\hbar}
\sqrt{2m (V_0 - E(0))}$ and $\Gamma_0 = \frac{\alpha_0 \hbar}{m}
\varrho(x_0;0).$

Using the property
$$
\frac{1}{y \sqrt{1 + y^2 }}= \frac{d}{dy} \ln\frac{y + \sqrt{y^2 +
1} - 1}{y + \sqrt{y^2 + 1} + 1}
$$
the integral (\ref{10}) is taken by parts and after that the slow
logarithmic function is taken out of integral. As a result we get
\begin{equation}\label{11}
- \frac{1}{2} \Gamma_0 t=  \left[ \ln\frac{y + \sqrt{y^2 + 1} - 1}{y +
\sqrt{y^2 + 1} + 1} - \ln\frac{y_0 + \sqrt{y_0^2 + 1} - 1}{y_0 +
\sqrt{y_0^2 + 1} + 1}\right] e^{\frac{1}{2}\alpha_0 x_0 y^2}.
\end{equation}

This equation can be solved iteratively. First we assume
that $y=y_0$ in the exponential function equation (\ref{11}) and get
\begin{equation}\label{12}
y_1(t) = \frac{2f(t;y_0)}{1 - f^2(t;y_0)}
\end{equation}
where $ f(t;y) = \nu(y)e^{- \frac{1}{2} \Gamma_0 \displaystyle
e^{-\alpha_0 x_0 y^2} t}$ and $ \nu(y) = \frac{y + \sqrt{1 + y^2}
- 1}{y + \sqrt{1 + y^2} + 1}, \ \ \mbox{or}\ \ y =
\frac{2\nu(y)}{1 - \nu(y)^2}$. At the next iteration we substitute
solution (\ref{12}) into the function $f(t;y)$ and repeat the
procedure,
\begin{equation}\label{14}
N(t) =
\frac{\hbar^2\alpha_0^2}{2m \widetilde\lambda}
\left(\frac{2f(t;y_1(t))}{1 - f^2(t;y_1(t))}  \right)^2
\end{equation}

The function $\nu$ in the above equations is always less than one.
$\nu$ tends to zero in the linear limit, when $\lambda \to 0$, and
one gets the exponential decay $N(t) = N_0  e^{-\Gamma_0 t}$. At a
large nonlinearity parameter $\lambda$, the function $\nu$ may be
close to one and the initial decay strongly deviates from the
exponential behavior. However, $\nu$ diminishes in the time course
and at large time the asymptotical behavior of the number of
particle $N(t)$ becomes exponential.

Before discussing the results it is worthwhile to estimate the
validity of the adiabatic approximation made while solving
equation (\ref{5}). For this we need to compare the time
derivative of the velocity $v_t$ near the exit point $x_0$,
neglected in the above calculations, with the right hand side of
equation (\ref{5}). The time dependence of the velocity results
mainly from the time dependence of the exit point, so that
\begin{equation}\label{15}
v_t = \frac{2\dot x_0}{mv(x_0)} V'_{ext}(x)|_{x=x_0} \approx
\frac{1}{mv(x_0)} \frac{V_0}{l}\dot x_0
\end{equation}
with $l$ being the typical length characterizing the potential
$V_{ext}(x)$. Since $x_0$ is the solution of equation $V_{ext}(x)
= E(0) - \widetilde \lambda N(t)$ we get $\dot x \approx
\frac{\dot N}{N} \frac{\widetilde\lambda N}{V_0} l.$ The right
hand side of equation (\ref{5}) can be estimated as $\approx
V_0/ml$ so that comparing it with the time derivative of the
velocity field (\ref{15}) we need to check the inequality
\begin{equation}\label{16}
\frac{v_tm l}{V_0} \approx \frac{\dot x_0}{v(x_0)} \approx
\tau_{tr} \frac{\dot N}{N}\frac{\widetilde \lambda N}{V_0} \ll 1 .
\end{equation}
Here $\tau_{tr} = l/v$ is so called traversal time, which roughly
corresponds to the time needed for a tunneling particle to
traverse the classically forbidden underbarrier region. It is
quite clear that at any reasonably high potential barrier the
tunneling rate is small enough so that $\tau_{tr}\dot N/N \ll 1$.
The ratio $\widetilde \lambda N /V_0$ cannot be large and in many
realistic cases it is even small at any reasonable number of particles
in the droplet. We may conclude that the inequality (\ref{16}) is
robust and holds at all reasonable parameters of the system, which
justifies the adiabatic approximation applied in this paper.

\begin{figure}
\includegraphics[width=9cm,angle=-90]{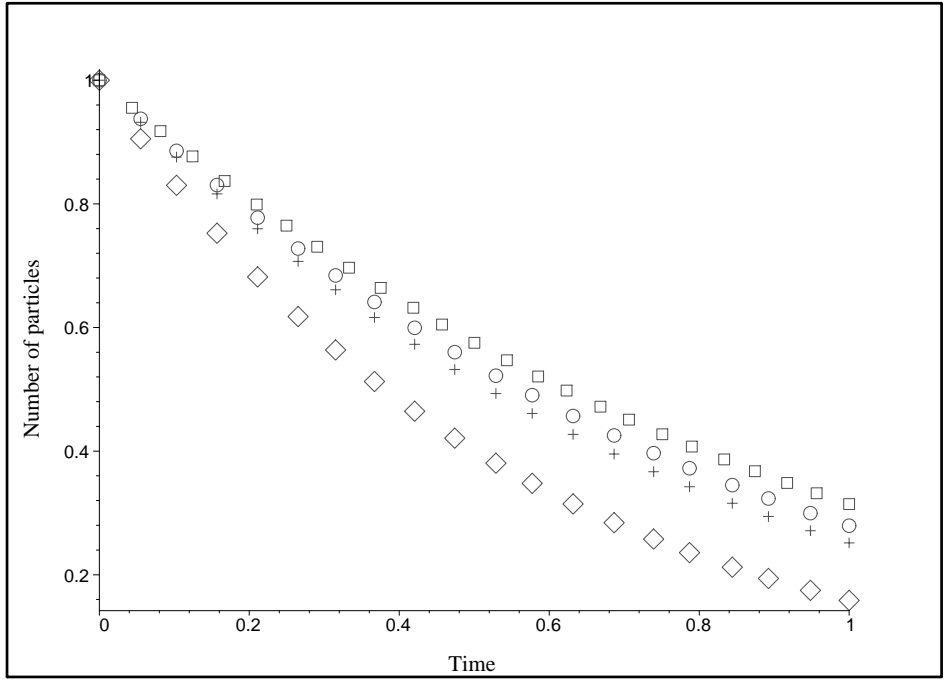}
\caption{Relative number of particles $\left(N(t)/N_0\right)$ as a function 
of time (in units $1/\Gamma$) for different values of the initial number of
particles $N_0/N_{GP}$ and $\alpha_0 x_0 = 3$: diamonds --- $N_0/N_{GP}  =0$;
crosses --- $N_0/N_{GP} = 0.004$; circles --- $N_0/N_{GP} = 0.04$; 
boxes --- $N_0/N_{GP} = 0.4$ } 
\label{f.4}
\end{figure}

Figure \ref{f.4} shows the time dependence of the number of
particle $N(t)$ in a droplet for different initial numbers $N_0$.
These numbers are scaled with the quantity $N_{GP}$ defined in
(\ref{10}), which can be thought of as a number of particles, at
which the nonlinear contribution to the single particle energy
$E(N)$ becomes comparable with the linear part $E(0)$. We see that
the time decay of the droplet is purely exponential, if the
initial number $N_0$ is small. It deviates from this exponential
behavior and becomes essentially slower at large initial numbers
of particles in the droplet. This deviation is stronger the
larger $N_0$ is, but at large time when the total number of
particles $N(t)$ becomes smaller the behavior of the decay curve
tends to the exponential one.

We may also find (directly from (\ref{10})) the initial slope of
the decay curve as a function of the initial number of particles
$N_0$,
\begin{equation}\label{17}
- \left.\frac{1}{N \Gamma_0 }\frac{dN}{dt}\right|_{t=0} =
  \sqrt{1 + \displaystyle 2\frac{N
}{N_{GP}} }\ \  e^{-\displaystyle \alpha_0 x_0 \frac{N}{N_{GP}}}.
\end{equation}
\begin{figure}
\includegraphics[width=9cm,angle=-90]{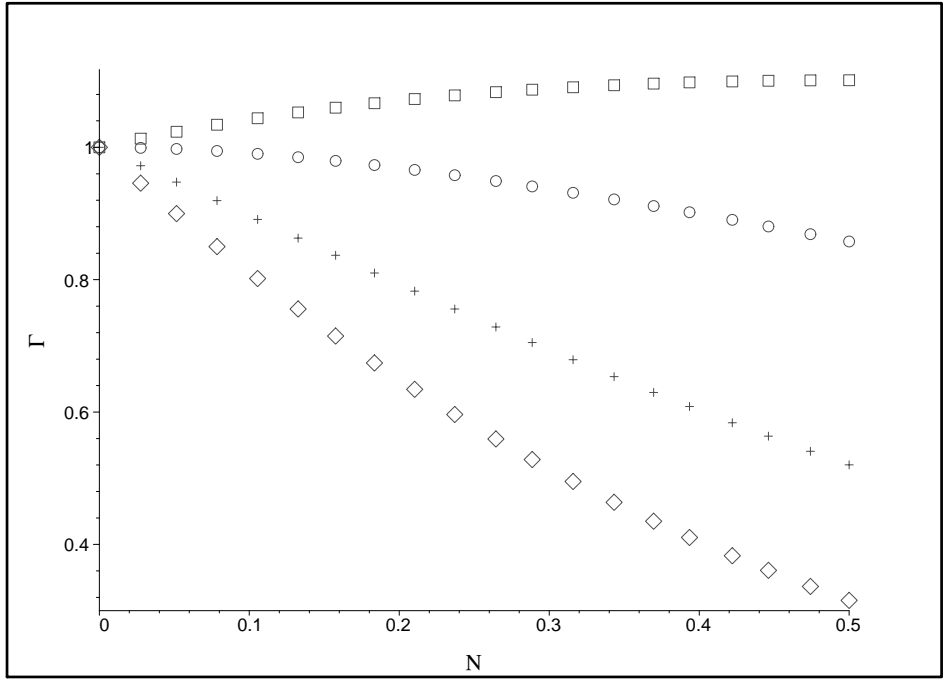}
\caption{Initial rate of evaporation $\Gamma$, in units of $\Gamma_0$ as
a function of the initial number of particles $N_0/ N_{GP}$ at for
several choices of $\alpha_0 x_0$: diamonds --- $\alpha_0 x_0 = 3$ ;
crosses --- $\alpha_0 x_0= 2$ ; circles
---  $\alpha_0 x_0 = 1$ ; boxes --- $\alpha_0 x_0 = 0.5$ } \label{f.5}
\end{figure}
It is interesting to note that the nonlinear effects usually cause
a decrease of the escape rate with the increasing initial number
of particles $N_0$ in the droplet as, e.g. is demonstrated in fig.
\ref{f.4}. On one hand, this trend is determined by the exponential
function in (\ref{17}), corresponding to the density of particles
at the exit point $x_0$, and the decrease is stronger for higher
and wider barriers, large $\alpha x_0$. On the other hand the
square root factor, i.e. the particle velocity at the exit
point, may result in an increase of the escape rate. The latter
may happen for rather low barriers when $\alpha_0 x_0 < 1$ (see
fig. \ref{f.5}).

We have demonstrated here how the hydrodynamic approach to the
description of a multiparticle quantum system lead to a solution
of the nonlinear problem of particle tunneling escape from a
trap. Using the adiabatic approximation (neglect of $v_t$),
justified for nonlinearities, which are not necessarily small, we
succeeded in obtaining an analytical solution in a one-dimensional
case for the time dependence of the number of trapped particles
and for the tunneling rates at different nonlinearity values. The
technique, we use, goes beyond the WKB approximation and, in
particular, divergencies at the turning point of the trap
potential do not appear. That is why applying the same technique
directly to three dimensional systems seems to be straightforward.
Contrary to the standard WKB approach velocity is not necessarily
a potential field (see, e.g. discussion in Ref. \onlinecite{lf})
and in three dimensional cases vortices can be considered.

{\bf Acknowledgment} Auhtors are indebted to S. Flach for discussions.


\begin{thebibliography}{99}

\bibitem{aemw95}   M.H.J. Anderson, J.R. Ensher, M.R. Matthews,
C.E. Wieman, Science {\bf 269}, 198 (1995).

\bibitem{dmaddkk95}  K.B. Davis, M.-O. Mewes, M.R. Andrews, N.J. van
Druten, D.S. Durfee, D.M. Kurn, and W. Ketterle, Phys. Rev. Lett.
{\bf 75}, 3969–3973 (1995).

\bibitem{bsth95} C. C. Bradley, C. A. Sackett, J. J. Tollett, and R. G. Hulet,
Phys. Rev. Lett. 75 , 1687 (1995).

\bibitem{fsec03} J.W. Fleischer, M. Segev, N.K. Efremidis, D.B.
Christodoulides, Nature {\bf 422},6928 (2003).

\bibitem{iasmsk98} S. Inouye, M. R. Andrews, J. Stenger, H. J. Miesner, D. M.
Stamper-Kurn, and W. Ketterle, Nature {\bf 392}, 151 (1998).

\bibitem {l9800} E. H. Lieb and J. Yngvason, Phys. Rev. Lett. 80 ,
2504 (1998); E. H. Lieb, R. Seiringer and J. Yngvason, Phys. Rev.
A 61 , 043602 (2000).

\bibitem{p61} L.P. Pitaevskii, Sov. Phys. JETP {\bf 13}, 451 (1961).

\bibitem{g63} E.P. Gross, J.Math.Phys. {\bf 4} , 195 (1963).

\bibitem{ss99}   C. Sulem and P. L. Sulem, The Nonlinear
Schr\"odinger Equation (Springer-Verlag, New York, 1999).

\bibitem{rp02} S.L. Rolston and W.D. Phillips, Nature {\bf 416},
219 (2002)

\bibitem{sspkpl04} Y. Shin, M. Saba, T.A. Pasquini, W. Ketterle, D.
E. Pritchard, and A.E. Leanhardt, Phys. Rev. Lett. {\bf 92},
050405 (2004).

\bibitem {wh99} J.E. Williams and M.J. Holand, Nature {\bf 568},
401 (1999).

\bibitem{mahhwc99} M.R. Matthews, B.P. Anderson, P.C. Haljan,
D.S. Hall, C.E. Wieman, and E.A. Cornell, Phys.Rev.Lett. {\bf 83},
2498 (1999).

\bibitem{mcwd99} K.W. Madison, F. Chevy, W. Wohlleben, and
J. Dalibard, Phys.Rev.Lett. {\bf 84}, 806 (1999)

\bibitem{igrcgglpk01} S. Inouye, S. Gupta, T. Rosenband, A.P.
Chikkatur, A. Gorlitz, T.L. Gustavson, A.E. Leanhardt, D.E.
Pritchard, and W. Ketterle, Phys.Rev.Lett. {\bf 87}, 080402
(2001).

\bibitem{lf} M. Levanda and V. Fleurov, Ann.Phys. NY {\bf 292}, 199 (2001);
ibid, J.Phys.: Condens Matter {\bf 6}, 7889 (1994)

\bibitem{itg02} M. Imamov-Tomasovic and A. Griffin, J.Low Temp.Phys.
{\bf 122}, 617 (2001)

\bibitem{foot} Single particle Wigner function may be obtained by 
integrating the coordinates and momenta of $N -1$ particles of an $N$ 
particle Wigner functions. A more detailed description can be found in 
\cite{lf}.

\bibitem{sw} A. Soffer and M.I. Weinstein, "Selection of the
Ground state...", Preprint 2001, to appear

\end{thebibliography}
\end{document}